\documentstyle[12pt,aasms4]{article} 
\newcommand{\keV}{ke\kern-.15emV}

\begin{document}
 
\title{EUV images of the Clusters of Galaxies A2199 and A1795: clear
evidence for a separate and luminous emission component}
 
\author{Richard~Lieu$\,^{1}$, Massimiliano Bonamente$\,^{1}$,
Jonathan~P.~D.~Mittaz$\,^{2}$}

\affil{\(^{\scriptstyle 1} \){Department of Physics, University of Alabama, 
Huntsville, AL 35899, U.S.A.}\\
\(^{\scriptstyle 2} \){Mullard Space Science Laboratory, UCL,
Holmbury St. Mary, Dorking, Surrey, RH5 6NT, U.K.}\\
}
 
\begin{abstract}
Since all of the five clusters of galaxies observed by the 
{\it Extreme Ultraviolet Explorer (EUVE)} deep survey telescope are found to
possess a diffuse EUV emitting component which is unrelated
to the hot intracluster medium (ICM)
at X-ray temperatures, the question concerning the
nature of this new component has been a subject of controversy.  Here we
present results of an EUV and soft X-ray spatial analysis of
the rich clusters Abell 2199 and 1795.  The EUV emission
does not
resemble the X-ray morphology of clusters:
at the cluster core the EUV contours are
organized; at larger radii they are
anisotropic, and are therefore unrelated to the hot ICM.
The ratio of EUV to soft X-ray intensity rises with
respect to cluster radius, to reach values $\sim$ 10 times higher than that
expected from the hot ICM.
The strong EUV excess which exists in the absence of soft X-ray excess
poses formidable problems to the non-thermal (inverse-Compton) scenario,
but may readily be explained as due to emission lines present only in the
EUV range.  In particular, warm gas produced by shock heating could account
for such lines without proliferation of bolometric luminosities and
mass budgets.
\end{abstract}

The `cluster soft excess' (CSE) phenomenon, which originated from
EUVE,
was confirmed by the
ROSAT and BeppoSAX
(Lieu et al 1996a,b;
Bowyer, Lampton and Lieu 1996; Fabian 1996; Mittaz, Lieu and Lockman 1998;
Bowyer, Lieu and Mittaz 1998; Kaastra 1998).  The non-thermal interpretation
of this phenomenon (Ensslin and Biermann 1998; Hwang 1997; Sarazin and 
Lieu 1998), favored on plausibility grounds over the original thermal
(warm gas) scenario, is recently supported by the
BeppoSAX and RXTE discoveries of hard X-ray tails in the spectra of 
several clusters (Kaastra 1998; Fusco-Femiano 
et al 1998; Rephaeli et al 1999), some of 
which exhibit similar radial trends as the CSE.
In this {\it Letter} we report first results on a spatial analysis of
the EUV which reveals clear and essential differences in
spatial morphology between the EUV and soft X-rays.  Our results concern
the rich clusters Abell 2199 and 1795; although a similar conclusion
exists also in the case of Virgo, A4038, and Coma, the analyses of these
will be presented in a subsequent and more detailed paper.

Abell 2199 was observed by the EUVE deep survey (DS) 
Lex/B (69 - 190 eV) filter for $\sim$
50,~000 sec.  Figure 1(a) is a radial profile of the cluster surface
brightness after removal of point source contributions and
subtraction of a background taken from the 
$>$ 15 arcmin region where there is no further evidence of cluster emission.
Standard correction procedures were applied\footnote{See `Deep Survey
Dead Spot Correction Algorithm', internal memo MMS/EUVE/0084/94, Center for
EUV Astrophysics, Berkeley}
to recover the signals lost
within a small region near boresight of $\sim$ 1 arcmin radius and
$\sim$ 3 arcmin away from the cluster center - a region known as the
deadspot, which was caused by the February 1993 observation of the intense
EUV source HZ 43.  This procedure ensures full integrity in the data reduction
procedure;
its actual effect is minor, resulting only 
in a slight, statistically insignificant enhancement of the
brightness in the central portion of the profile.

To compare with the X-ray behavior, data from 
a {\it ROSAT} Position Sensitive
Proportional Counter pointing which took place in
July 1990 were extracted from the
archive\footnote{HEASARC, available at
http://heasarc.gsfc.nasa.gov/docs/rosat/archive.html, and maintained by
the ROSAT/ASCA Guest Observer Facility.}.  Again point sources were
removed, and a background as determined from the outermost annulus 
was subtracted after vignetting corrections.  A radial profile of the
EUV to soft X-ray ratio is shown in Figure 1(b).
To perform a quantitative assessment of the trend,
we also show in Figure 1(b) the expected value of the ratio if the
emission originated from the virial gas alone.  This expected value is
obtained by modeling the PSPC 0.2 - 2.0 keV 
(PH channels 18 - 200) data of concentric annuli with a single
temperature MEKAL thin plasma code (Mewe, Gronenschild \& van den Oord 1985;
Mewe, Lemen \& van den Oord 1986; Kaastra 1992), using an abundance of
0.5 solar (David et al 1993) and
line-of-sight column density as
measured by a dedicated observation at Green Bank (for further details
see Lieu et al 1996a,b), which reported $N_H =$ 8.3 $\times$ 10$^{19}$
cm$^{-2}$ with a nominal error of $<$ 10$^{19}$ cm$^{-2}$.
No CSE was evident in the PSPC data: specifically the
model satisfactorily accounts for all the data of the employed
PH channels.

However, the Lex/B data indicate
substantial CSE, as it can be seen from Figure 1b
that the relative strength of 
the Lex/B CSE rises with radius, analogous to Abell 1795 (Mittaz, Lieu
\& Lockman 1998). This is borne-out even more by the 2-D images below.
We emphasize that the effect of concern is only slightly
modified if more accurate abundances (Mushotzky et al 1996)
were used to compute the expected
softness ratio - 
the basic notion of a rising trend in this ratio does
{\it not} appear to be an abundance issue.
Given 
that most of the extragalactic EUV radiation is absorbed, and that the DS data
are background limited, the radially rising softness ratio suggests
a possible spatial extent of the EUV emission far larger than
our current measurements.

To investigate the spatial distribution of the EUV, we adaptively smoothed the
DS event image with a gaussian filter which encloses a S/N (signal-to-noise
ratio\footnote{Since EUVE data are background limited,
this S/N is calculated as the ratio $(S-B)/\sqrt{S}$, where $S$ is the
number of counts enclosed by the filter and $B$ is the detector background
estimated from regions free of cluster emission.})
of $\geq$ 5 $\sigma$.  In the central region of
the cluster where the signals are strong, a minimum filter size equal to
that of the 
DS point spread
function (PSF, Sirk et al 1997) was enforced.
Contours of the 
smoothed and background subtracted surface brightness are
shown in Figure 2(a).
The lowest contour level plotted is
$\sim$ 7.5 \% above background; since no part of the entire 
2$^o$ $\times$ 0.5$^o$ area of the Lex/B filter
has such a high brightness, other than the cluster region as shown,
the detected emission
is not an apparent effect caused by random background variations. 

To further establish the reality of the extended component we produced
similarly smoothed images of (a) two 
blank field pointings; and (b) an observation
of the globular cluster X-ray source M15 (for 1 -- 6.5 keV detection
see Callanan et al 1987), which showed that the source was visible in the EUV.
The uniformity of the background
in (a), especially the absence of any enhancements over a large area
centered at boresight - an area
where the observed cluster targets normally occupy, excludes the
possibility
that diffuse cluster EUV is a systematic
detector spatial effect.  In (b), the lack of any extended
halo around M15, despite the
source having a peak surface brightness comparable
to that of the two clusters,
also rules out
leakage of central source radiation in the EUV or X-rays as origin of the
cluster syndrome - a conclusion fully consistent with our understanding
of the DS PSF.  All the image data referred to in this {\it Letter} are
available as FITS files\footnote{These can be downloaded via anonymous 
ftp from the address cspar.uah.edu/input/max/.}.

For comparison with spatial behavior in the X-rays, we show in Figure 2(b)
a background subtracted and exposure corrected
soft X-ray contour map of A2199, using the R2 band
data of a ROSAT
PSPC observation (R2 refers to PH channel 18-41, or
$\sim$ 0.2 -- 0.3 keV by channel boundaries) and a gaussian
smoothing filter commensurate in size with the PSPC PSF in this passband
(note that S/N is not a problem for the PSPC data).
Despite the proximity of the
R2 and Lex/B passbands, resemblance between
Figures 2(a) and 2(b) is only restricted to the
cluster core, where both images show a correlating set of
organized contours.
At larger radii the R2 contours remain regular and symmetrical, reminiscent
of emission from a hydrostatic gas, but the Lex/B contours
do not share this property.  

To assess the statistical significance of
the Lex/B anisotropies,
we divided the image into concentric annuli moving outwards from
the cluster emission centroid.
The surface brightness of each annulus is
computed over octants, and a $\chi^2$ test is applied to
the hypothesis that the octant counts are all consistent with their
azimuthally averaged value.  In Table 1 we show $\chi^2_{red}$ for four equally
spaced annuli between 0 and 12 arcmin radius.  It can be seen that for
regions outside 6 arcmin radius emission anisotropy is asserted with
a confidence level of $>$ 98.5 \% (corresponding to $\chi^2_{red} \geq$ 2.4 for
8 degrees of freedom).  We repeated this test on blank field data of
comparable exposure, using
the same near-boresight area of the DS detector where cluster observations
were made.  Results, also shown in Table 1, indicate that the detector
background distribution does not have similar anomalies.
Furthermore, detailed plots (not shown)
reveal that the A2199 anisotropy is due to deviations of the data
from the average value in directions where spatially asymmetric features
are apparent in Figure 2(a).  Visual comparison with the optical
map of the Digital Sky Survey revealed no obvious anisotropy in the
density and brightness
distribution of member galaxies within the relevant region.

From the radial trend of the
Lex/B : R2 ratio (Figure 1b) it was evident that the relationship between
the EUV and the hot ICM decrease
with radius.  This is borne-out more strongly in 2-D by Figure 2(c), which
indicates an even sharper radial increase of the ratio
in directions normal to the Lex/B brightness contours.
The lack of any resemblance
between Figures 2(a) and 2(b) at the outer regions of the cluster
provides crucial further confirmation
that the EUV emitting phase leads an existence entirely
separate from the hot ICM.  In particular the possibility of interpreting
the CSE as due to
our overestimate of the line-of-sight Galactic absorption
of the hot ICM radiation (e.g. Arabadjis \& Bregman 1998)
is completely excluded by these latest data.

Apart from A2199, in this {\it Letter} we also report results of a similar
study of the rich cluster Abell 1795.  An earlier paper (Mittaz, Lieu
\& Lockman 1998) showed radial profiles of both the EUV surface brightness and
the emission softness ratio for this cluster, it also
showed a feeble CSE in the PSPC data - such information will not
be repeated.  Here we focus on a 2-D image analysis, where all the 
foregoing discussions regarding adopted techniques apply.
As before, we now show
in Figure 3 contours of the DS Lex/B and PSPC R2-band surface brightness,
and of the Lex/B : R2 ratio.  The maps lead us to draw conclusions
about A1795 similar to those of A2199, viz. that the Lex/B  contours are
organized (but differently shaped from the R2 ones) at the core, are
anisotropic at larger radii (see also Table 1) with no similar effects
in the optical,
and that the rising radial trend of softness ratio
is in agreement with that reported
in our earlier paper.

Several interesting physical deductions can be made from a spatial
comparison between the EUV and X-rays.  The anisotropic EUV emission could,
within the context of the
inverse-Compton scenario, reflect the
distribution of the intracluster magnetic field strength (Ensslin, Lieu
\& Biermann 1999) in that electrons in regions with $>$ a few $\mu$G
fields would have been removed by synchrotron losses as
diffusive replenishment is always negligibly slow
(V\"olk, Aharonian \& Breitschwerdt 1996).
On the other
hand, if the emission is thermal,
as originally advocated, this would imply the existence of warm gas
which, by virtue of its lack of hydrostatic equilibrium at these lower
temperatures, could have
an anisotropic (and possibly
clumped) distribution as well.

Why do the Lex/B and R2 maps appear so different, given that their
energy passbands have substantial overlap ?  
Our simulations of the instrumental performances
indicate that
this could happen only if
the EUV component has a spectrum which does not exceed $\sim$ 0.2 keV.
For a non-thermal origin the 
behavior implies that the bulk of the relativistic electrons have energies
below 200 MeV, a cut-off effect which is most obviously
understood as due to aging.  
However,
a major difficulty concerns the
strong Lex/B excess accompanied by
little or no R2 excess, as this requires an age of $\ge$ 3 Gyr and hence 
a very significant spectral evolution (i.e. losses).  By the time
the spectrum at the present epoch explains the observations, at
injection the
electron pressure alone would have far exceeded that of the hot ICM gas,
leading to an unreasonably short lifetime estimate for the gas.

It is more natural to interpret the marked contrast in the
behavior of the Lex/B and R2 bands as signatures of bright emission lines
which are present only in the EUV range.  Although exotic line species
arising from an interaction process which may involve dark matter cannot
totally be excluded, a specific scenario of this kind is not available.  
However,
lines could simply be a direct witness to the presence of
thermal gas.  For example,
warm under-ionized gas shocked in mixing layers around
cool clouds (Fabian 1997)
{\it can} produce a blend of 
intense lines exclusively
within the Lex/B energy range, thereby limiting the bolometric luminosity
and (hence) the gas mass budget and mass cooling rate to more reasonable values.
The rise in relative EUV excess with radius
may then be a density scaling effect - at
larger radii the rarer gas stays longer in this out of equilibrium phase.
Detailed hydrodynamic simulation of the thermal model, with associated mass
implications, is work in progress.

We thank an anonymous referee for helpful comments.

\newpage

\begin{table}[h]
\begin{center}
\begin{tabular}{cccc} 
 Annulus (arcmin) & A2199 & A1795 & Blank field \\ \hline
  0-3  & 1.25 &1.53& 0.64 \\ 
  3-6  & 1.6&2.2 & 0.13 \\ 
  6-9  & 2.4 &2.5& 0.4 \\ 
 9-12  & 2.62 &4.7&1.1 \\ \hline
    
\end{tabular}
\end{center}
\caption{Reduced $\chi ^2$ goodness-of-fit test of the azimuthal symmetry in 
the Lex/B surface
brightness distribution around several 
concentric annuli.  Each annulus was divided into
8 equal octants and the $\chi ^2_{red}$ value was computed for a model of
of constant brightness given by averaging over all angles.  The center
of these annuli is either coincident with the EUV centroid of the
relevant cluster or, in the case of the background observation, with a
near-boresight
position of the DS detector which was typically used for pointing at
cluster centroids.  Each octant has at least a few $\times$ 100 counts
to ensure validity of the $\chi^2$ statistic.}
\label{tab1}
\end{table}

\newpage

\noindent{\bf Figure Captions}
 
\noindent Figure 1a.  Radial profile of the EUV
surface brightness of A2199.  A background as determined from the data
taken beyond the radius of 15 arcmin was subtracted.  \\

\noindent Figure 1b. Radial profile of the DS Lex/B to PSPC R2 band
ratio.  The dotted line represents the expected band ratio of
$\sim$ 0.03 from a
single temperature fit to the DS and PSPC data, using the MEKAL code.  The
model for line-of-sight Galactic absorption is
Balucinska-Church \& McCammon (1992); if 
the Morrison \& McCammon (1983) model was used instead, the predicted ratio
would become 0.037.  The EUV depletion within 2 arcmin radius can be
explained by the addition of an intrinsic absorbing column of
$\sim$ 3.5 $\times$ 10$^{19}$ cm$^{-2}$.  Elsewhere the rise of the ratio
(and hence the degree of soft excess)
with radius is evident.  \\

\noindent Figure 2a.  
DS Lex/B contours of the surface brightness of Abell 2199 (for
details see text).
The peak emission is
$1.98 \times 10^{-3} ph/arcmin^2/s$, and 
contours are labeled as percent of this peak value, with
the lowest contour at $8.4 \times 10^{-5} ph/arcmin^2/s$. \\

\noindent Figure 2b.
PSPC R2 band contours of the surface brightness of Abell 2199 (for details
see text).  The
peak emission is $8.3 \times 10^{-2} ph/arcmin^2/s$.  Next contours are $1.66
\times 10^{-3}, 2.8 \times 10^{-3}, 4.2 \times 10^{-3}, 7.2 \times 10^{-3}, 1.4
\times 10^{-2}, 2.9 \times 10^{-2}$, 
and $7.4 \times 10^{-2} ph/arcmin^2/s $. \\

\noindent Figure 2c.  Contours of DS Lex/B to PSPC R2 band ratio of Abell 2199. 
The dotted
lines mark regions where the DS brightness is 6 \% above the
background, and the contours are labeled in units of 1 \%.  The expected
value of this ratio was given earlier (see caption of Figure 1b). \\

\noindent Figure 3a.
DS Lex/B contours of the surface brightness of Abell 1795 (for
details see text).
The peak emission is
$3 \times 10^{-3} ph/arcmin^2/s$, and
contours are labeled as percent of this peak value, with 
the lowest contour at
$1.6 \times 10^{-4} ph/arcmin^2/s$ (5 \% above background). \\

\noindent Figure 3b.  
PSPC R2 band contours of the surface brightness of Abell 1795 (for details
see text).  The  
peak emission is $9.5 \times 10^{-2} ph/arcmin^2/s$. Next contours are $8.6 \times
10^{-4}, 1.3 \times 10^{-3}, 2.1 \times 10^{-3}, 3.6 \times 10^{-3},6.7 \times  10^{-3}, 9.8 \times 10^{-3},  1.4 \times 10^{-2} ,3 \times 10^{-2}$ and $4.5 
\times 10^{-2} ph/arcmin^2/s$. \\

\noindent Figure 3c.
Contours of DS Lex/B to PSPC R2 band ratio of Abell 1795. The dotted
lines mark regions where the DS brightness is 25 \% above the
background, and the contours are labeled in units of 1 \%.  The expected
value of this ratio, based on a single temperature ICM, is
1.8 \% (Mittaz, Lieu, \& Lockman 1998). \\

\vspace{2mm}

\noindent
{\bf References}

\noindent
Arabadjis, J.S., Bregman, J.N., 1998, {\it astro-ph/9810377}. \\
\noindent 
~Balucinska-Church, M. and McCammon, D., 1992, {\it
Astrophys. J.} {\bf 400}, \\
\indent 699--700 \\
\noindent
~Bowyer, S., Lampton, M., Lieu, R. 1996, {\it Science}, {\bf 274}, 1338--
1340. \\ 
\noindent
~Bowyer, S., Lieu, R., Mittaz, J.P.D. 1998,
{\it The Hot Universe: Proc. 188th \\
\indent IAU Symp., Dordrecht-Kluwer}, 52. \\
\noindent
~Callanan, P.,J., Fabian, A.,C., Tennant, A., F., Redfern, R., M. and Shafer,\\
\indent R., A. 1987, {\it M.N.R.A.S.}, {\bf224}, 781--789.\\
\noindent
~David, L. P., Slyz, A., Jones, C., Forman, W. and Vrtilek, S.D. 1993,\\
\indent {\it Astrophys. J.}, {\bf 412}, 479--488.\\ 
\noindent
~Ensslin, T.A., Biermann, P.L. 1998, {\it Astron. Astrophys.}, {\bf 330},
90--98. \\
\noindent
~Ensslin, T.A., Lieu, R. and Biermann, P.L., {\it submitted to Astron. \\
\indent Astrophys.}, astro-ph 9808139. \\
\noindent
~Fabian, A.C. 1996, {\it Science}, {\bf 271}, 1244--1245. \\
\noindent 
~Fabian, A.C. 1997, {\it Science}, {\bf 275}, 48--49. \\
\noindent
~Fusco-Femiano, R., Dal Fiume, D., Feretti, L., Giovannini, G., Matt, G.,\\
\indent Molendi, S. 1998, {\it Proc. of the 32nd COSPAR Scientific Assembly, \\
\indent Nagoya, Japan (astro-ph 9808012)}. \\
\noindent 
~Hwang, C. -Y. 1997, {\it Science}, {\bf 278}, 1917--1919. \\
\noindent
~Kaastra, J.S. 1998, {\it Proc. of the 32nd COSPAR Scientific Assembly, \\
\indent Nagoya, Japan }. \\
\noindent
~Kaastra, J.S. 1992 in \it An X-Ray Spectral Code for OpticallyThin Plasmas \rm \\\indent
(Internal SRON-Leiden Report, updated version 2.0) \\
\noindent
~Lieu, R., Mittaz, J.P.D., Bowyer, S., Lockman, F.J.,
Hwang, C. -Y., Schmitt, \\ 
\indent  J.H.M.M. 1996a, \it Astrophys. J.\rm, {\bf 458}, L5--7. \\
~Lieu, R., Mittaz, J.P.D., Bowyer, S., Breen, J.O.,
Lockman, F.J., \\
\indent Murphy, E.M. \& Hwang, C. -Y. 1996b, {\it Science}, {\bf 274},
1335--1338. \\
\noindent
~Lieu, R., Ip W.-I., Axford, W.I. and Bonamente, M. 1999, {\it ApJL} , \\
\indent {\bf 510}, 25--28.\\
\noindent 
~Mewe, R., Gronenschild, E.H.B.M., and van den Oord, G.H.J., 1985 \\
\indent {\it Astr. Astrophys. Supp.}, {\bf 62}, 197--254  \\ 
\noindent Mewe, R., Lemen, J.R., and van den Oord, G.H.J. 1986, 
\it Astr. Astrophys. Supp.\rm, \\
\indent {\bf 65} 511--536 \\
\noindent
~Mittaz, J.P.D., Lieu, R., Lockman, F.J. 1998, {\it Astrophys. J.}, {\bf 
498},
L17--20. \\
\noindent
~Morrison, R. and McCammon D., 1983, {\it Astroph. J.}, {\bf 270}, 119--122.\\
\noindent 
~Mushotzky, R., Loewenstein, M., Arnaud, K.A., Tamura, T., Fukazawa, Y.,
Matsushita, K.,\\
\indent Kikuchi, K., \& Hatsukade, I., 1996, {\it Astrophys. J.}, {\bf 466},
686. \\
\noindent
~Rephaeli, Y., Gruber, D. and Blanco, P. 1999, {\it ApJL}, {\bf 511}, 21 \\
\noindent
~Sarazin, C.L., Lieu, R. 1998, {\it Astrophys. J.}, {\bf 494}, L177--180. \\
\noindent Sirk, M.M. et al 1997, {\it Astrophys. J. Supp.}, {\bf 110},
347 . \\
\noindent
~V\"olk, H.J., Aharonian, F.A., Breitschwerdt, D. 1996, {\it Space Sci.
Rev.}, {\bf 75},

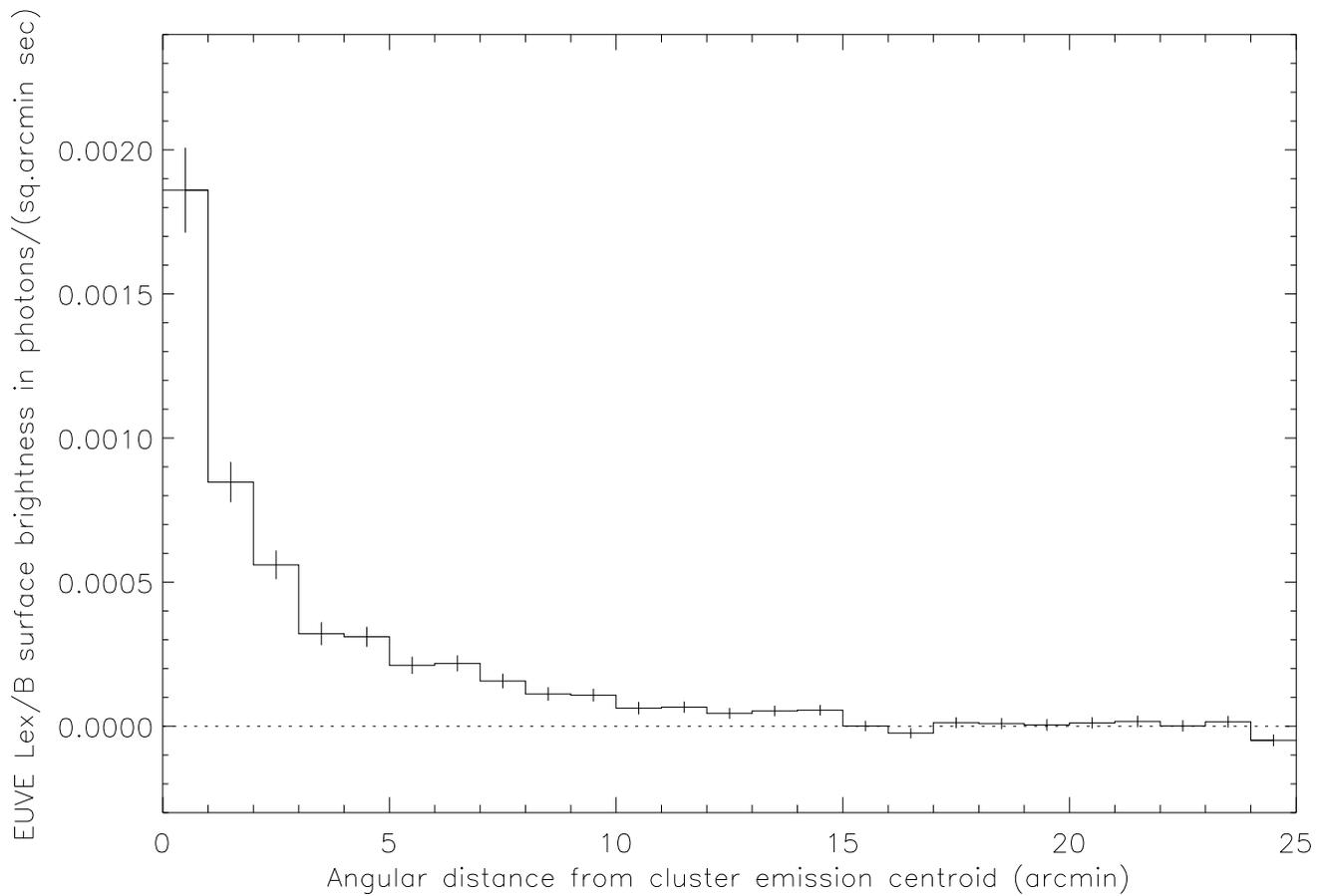
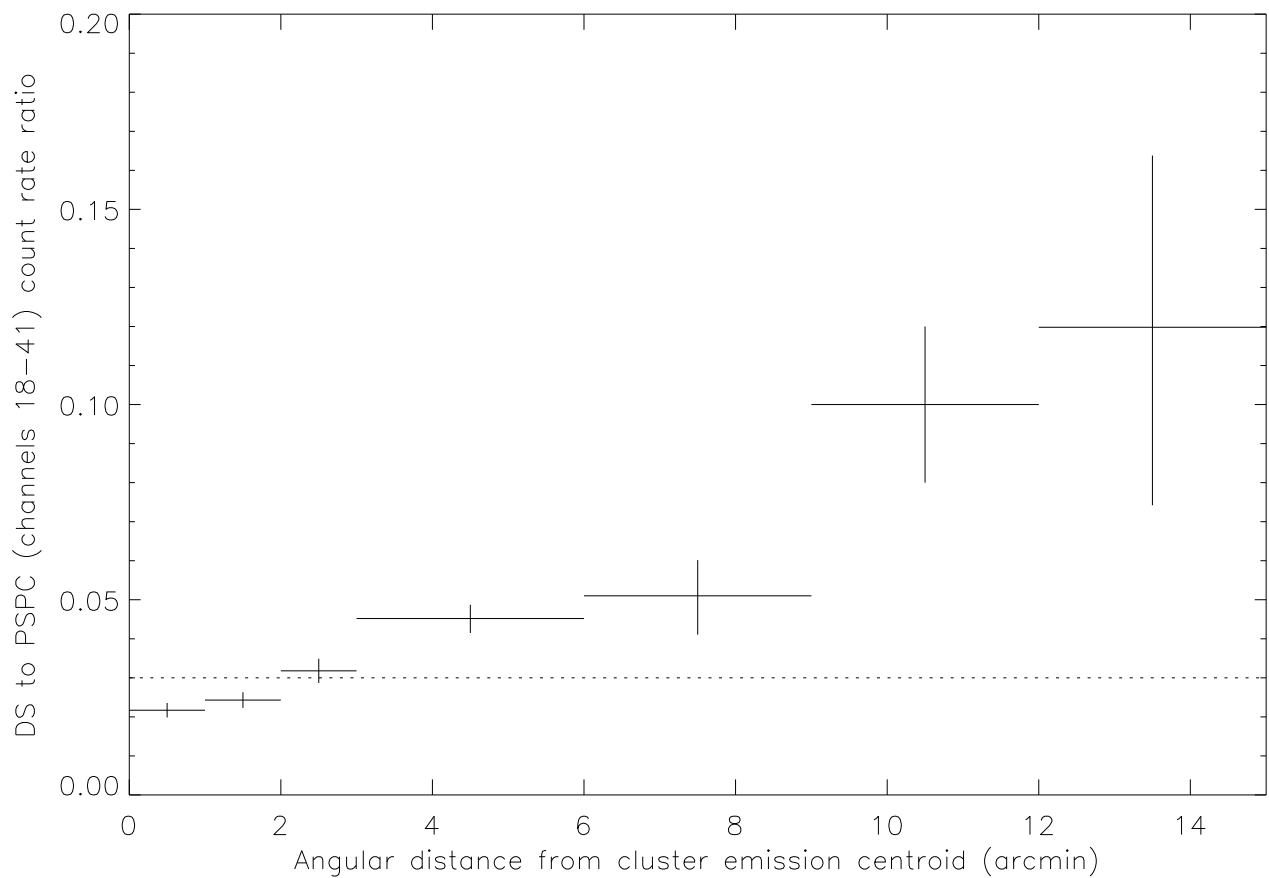
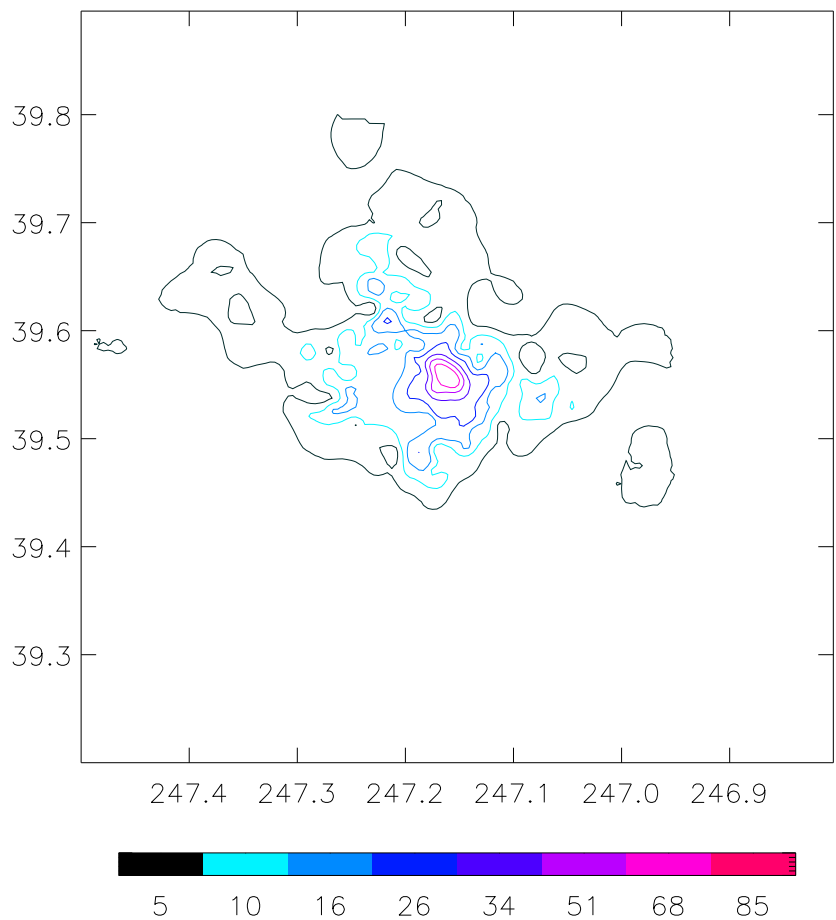
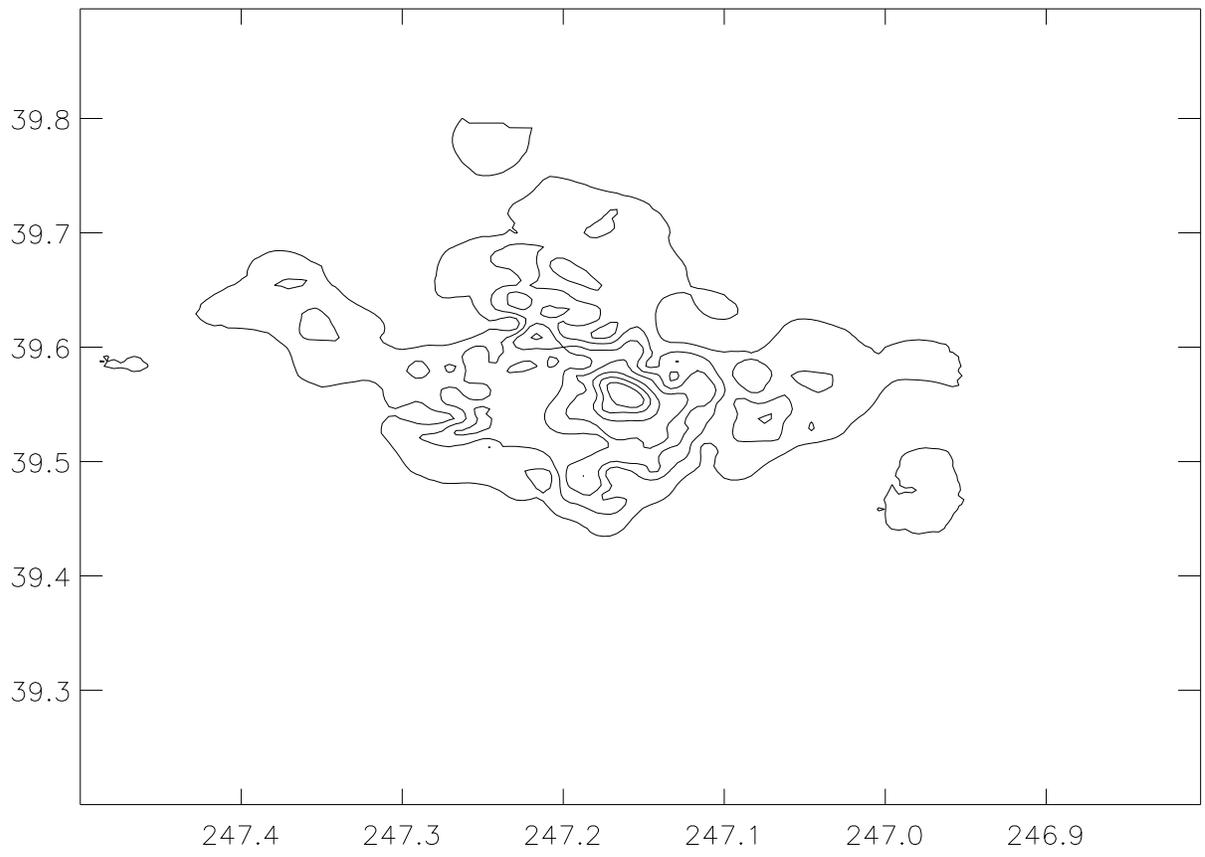
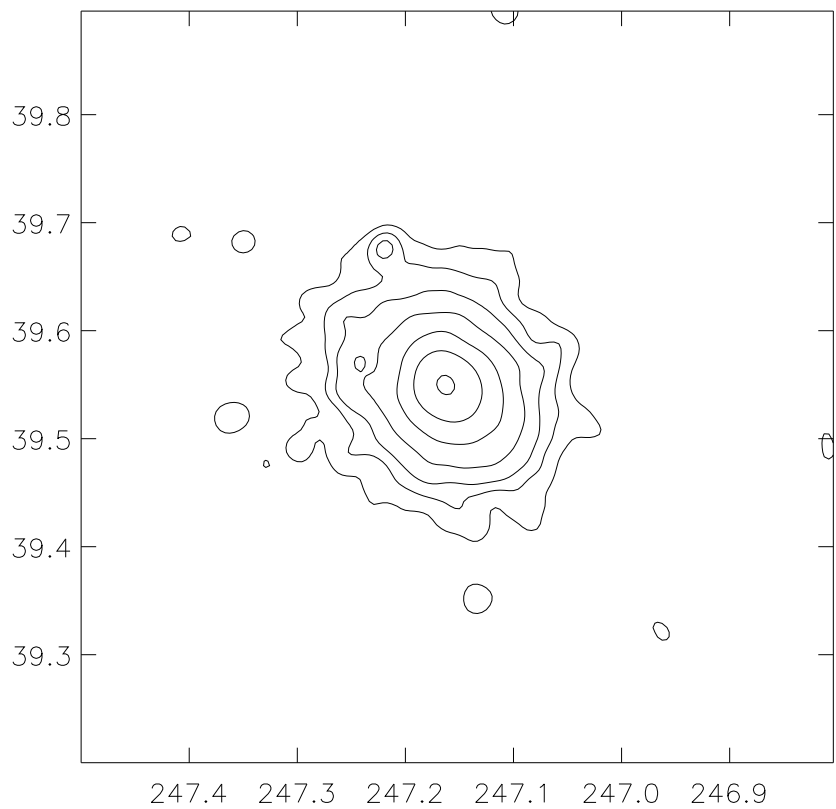
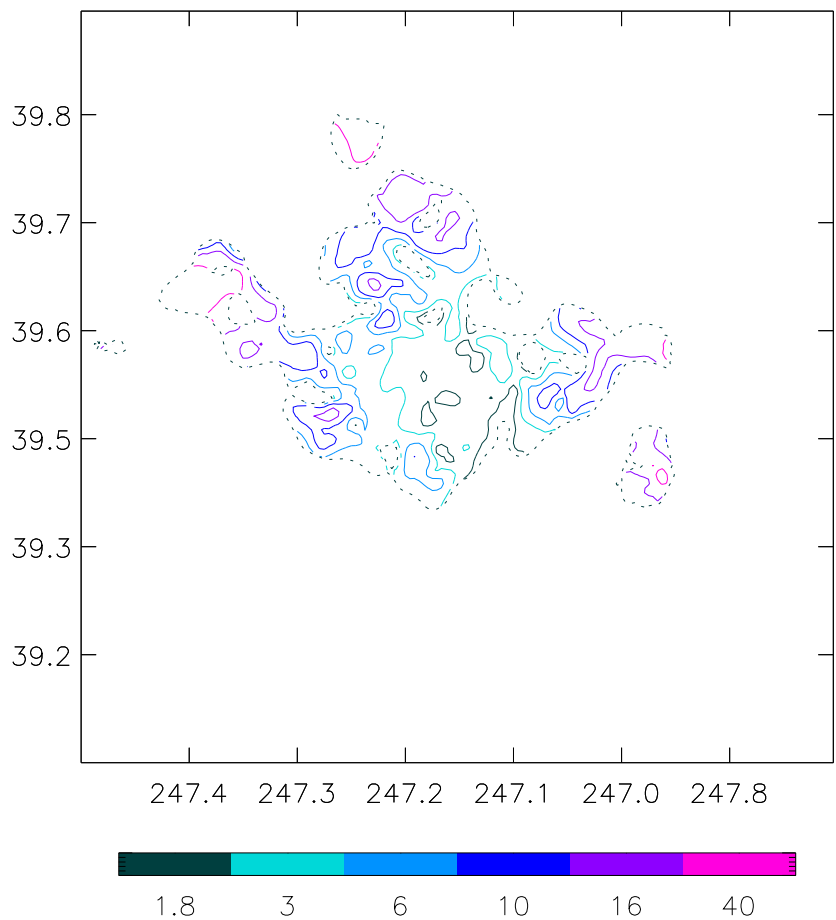
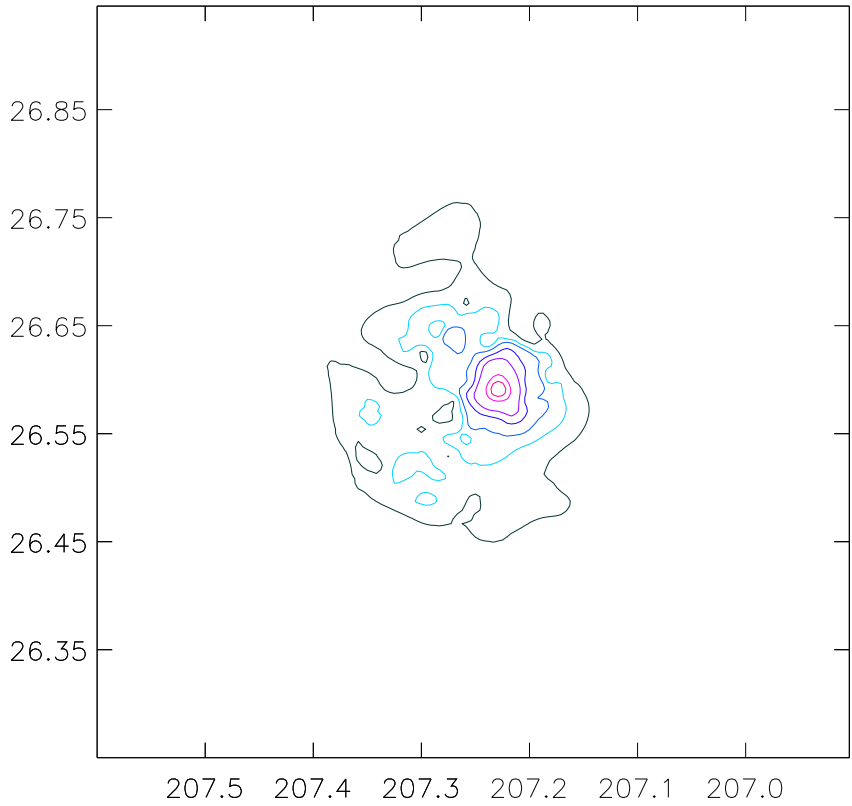
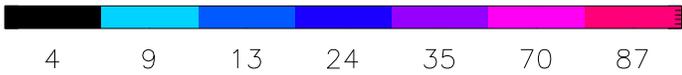
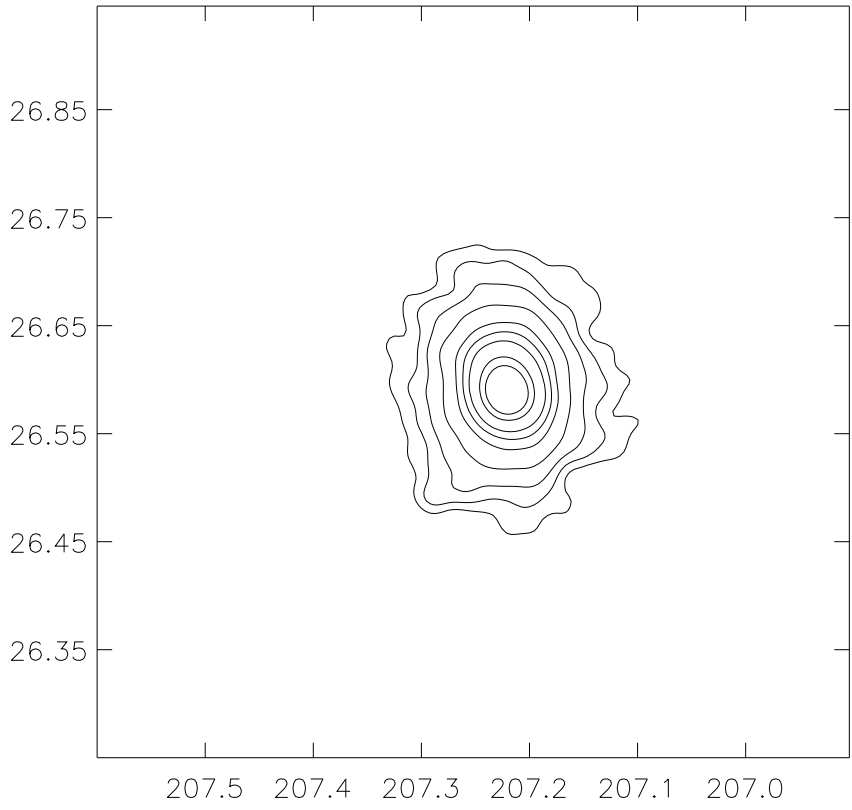
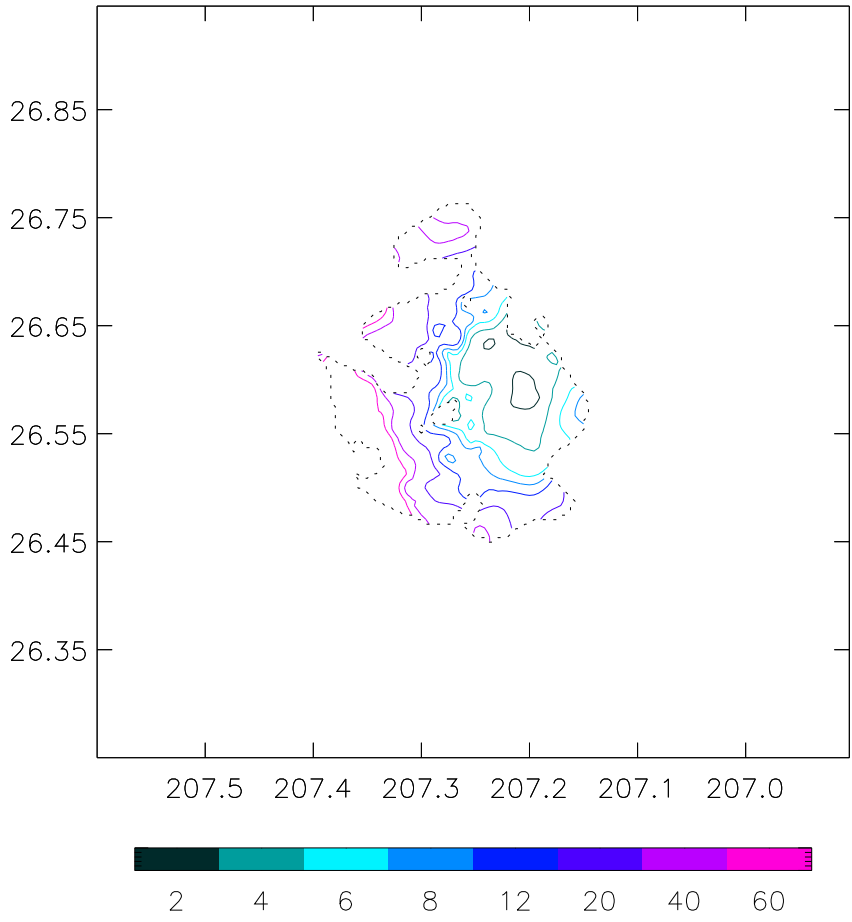
\end{document}